\documentclass[aps,prl,superscriptaddress,twocolumn,showpacs]{revtex4-1}

\usepackage[T2A]{fontenc}
\usepackage[cp1251]{inputenc}
\usepackage{graphicx}
\usepackage{psfig}
\usepackage{amsmath}
\usepackage{caption2}
\usepackage{floatflt}
\usepackage{dsfont}
\usepackage{mathrsfs}
\DeclareMathOperator\arctanh{arctanh}
\begin{document}
\title{Inverted pendulum state of a polariton Rabi oscillator}
\author{\firstname{N.~S.}~\surname{Voronova}}
\email{nsvoronova@mephi.ru}
\affiliation{%
National Research Nuclear University MEPhI (Moscow Engineering Physics Institute), 115409 Moscow, Russia}
\affiliation{Russian Quantum Center, 143025 Skolkovo, Moscow region, Russia}
\author{\firstname{A.~A.}~\surname{Elistratov}}
\affiliation{ Institute for Nanotechnology in Microelectronics RAS, 119334 Moscow, Russia}
\author{\firstname{Yu.~E.}~\surname{Lozovik}}
\affiliation{Institute for Spectroscopy RAS, 142190 Troitsk, Moscow, Russia}
\affiliation{Moscow Institute of Physics and Technology (State University), 141700 Dolgoprudny, Moscow region, Russia}
\affiliation{Moscow Institute of Electronics and Mathematics, HSE, 101000 Moscow, Russia}

\begin{abstract}
Exciton-photon beats known as polariton Rabi oscillations in semiconductor microcavities are usually excited by short pulses of light. We consider a different pumping scheme, assuming a cw pumping of the Rabi oscillator from an exciton reservoir. We account for the initial pulse of light setting the phase, exciton decay due to exciton-phonon and exciton-exciton scattering, photon leakage, and blueshift of the exciton resonance due to interactions. We find non-trivial stationary solutions reminiscent of Kapitza pendulum, where polaritons are accumulated at the upper branch while the lower branch empties.
\end{abstract}
\pacs{67.85.Hj, 03.75.Mn, 71.36.+c}

\maketitle

Light composite quasiparticles that occur due to strong coupling of quantum well excitons with photons confined in a semiconductor microcavity --- exciton polaritons --- have shown the ability to display macroscopic quantum coherence, including Bose-Einstein condensation \cite{kasprzak,balili}, superfluidity \cite{carusotto,amo1,amo2,amo3}, and varieties of bosonic Josephson phenomena \cite{wouters,sarchi,shelykh,read,konstantinos,abbarchi,voronova_prl}.
The underlying principle of the polariton physics is that of Rabi oscillations, which are considered a signature of the strong exciton-photon coupling regime in the microcavity. From the point of view of classical optics they can be viewed as the effect of interference of two coherent electromagnetic waves emitted at different frequencies corresponding to the lower (LP) and upper (UP) polariton branches. As a result, the intensity of light emitted from the cavity oscillates with a terahertz frequency corresponding to the splitting between upper and lower polariton frequencies, which has been recently observed with high precision \cite{ultrafast}.
It is important that also the excitonic population in the system oscillates in time. The excitonic oscillations
having the same frequency but opposite phase compared to photonic oscillations can be measured
independently e.g. by the pump probe Kerr rotation technique \cite{masha}.

As both excitons and cavity photons are bosons, polariton Rabi oscillations may be considered as beats in a system of two coupled harmonic oscillators. One of these oscillators is essentially non-linear:
repulsive exciton-exciton interactions result in the time-dependent blueshift of the exciton energy.
This blueshift contributes to the detuning between exciton and photon modes in microcavities.
Since it oscillates with the Rabi frequency, the non-linear Rabi oscillator can be viewed as a parametric oscillator.

Another important feature of the polariton system is its driven and dissipative character. Due to the finite quality factor of any realistic microcavity, cavity photons may tunnel through the Bragg mirrors --- as an advantage, allowing the condensate properties to be accessed for measurement. Excitons, too, may escape from the coherent Rabi oscillator due to scattering with acoustic phonons and other excitons. This leakage of photons and excitons may be compensated by pumping in the optical experiment.
Resonant optical pumping creates photons in the system, that may be converted into excitons due to the Rabi oscillations. In addition, a non-resonant optical pumping \cite{kasprzak} or electrical injection \cite{sven_el} are capable of creating an excitonic reservoir that would pump excitons into the coherent Rabi oscillator.

It has been argued recently that stimulated exciton pumping may bring the Rabi oscillator to a PT-symmetric state characterized by permanent Rabi oscillations \cite{rabi-permanent}. In this work, we study the effect of exciton pumping further and demonstrate that it may lead to the appearance of a non-trivial stationary state, where the upper polariton branch is strongly occupied with exciton-polaritons while the lower polariton branch is essentially empty. This state is sustained despite the strong leakage of excitons from the upper polariton branch due to the acoustic phonon assisted scattering and exciton-exciton scattering.
We draw an analogy between this peculiar solution of the non-linear driven-dissipative Rabi problem and
classical Kapitza pendulum. We discuss the critical conditions of excitation of this ``inverted pendulum'' state in realistic microcavities.

\begin{figure}[t]
\renewcommand{\captionlabeldelim}{.}
\includegraphics[width=\columnwidth]{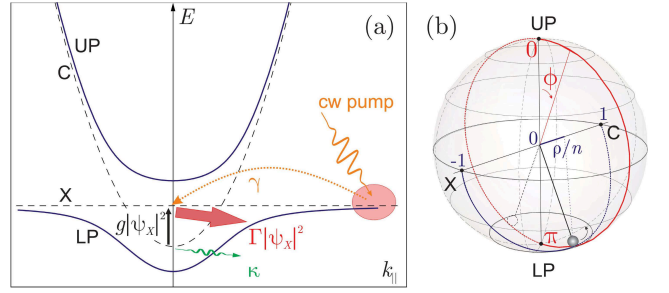}
\caption{\small (color online) (a) The LP and UP branches, and the photon and exciton dispersions, separated by the oscillating blueshift $g|\psi_X|^2$. The exciton subsystem is being fed linearly from the reservoir with the rate $\gamma$ and decays through non-linear scattering processes with the rate $\Gamma |\psi_X|^2$, while the photon counterpart decays linearly with the rate $\kappa$. (b) The rotated Bloch sphere representation merged with the pendulum analogy, with the relative phase $\phi$ and the normalized population imbalance $\rho/n$ changing as marked.}
\label{fig_1}
\end{figure}

It is convenient to describe the polariton Rabi oscillator in the exciton-photon basis, adopting the complex Ginzburg-Landau model of Refs.~\cite{borgh,wouters1,haug}.
We are interested in the temporal evolution of the system, and will consider the homogeneous case when there is no external trapping potential and the pumping is spatially uniform. Assuming zero wavevector, we therefore omit all the spatial derivatives. Taking into account that the pumping of polaritons is going through excitons and the natural decay is governed by the photon leak out of the cavity, this model reads in dimensionless form as follows:
\begin{equation}\label{cGL_1}
\begin{split}
i\partial_t\psi_C &= \!\Bigl[\epsilon^0_C - i\kappa\Bigr]\psi_C +\! \frac{1}{2}\,\psi_X, 
\\
i\partial_t\psi_X &= \!\Bigl[\epsilon^0_X + g|\psi_X|^2 \!\!+ i(\gamma-\!\Gamma|\psi_X|^2)\Bigr]\psi_X +\! \frac{1}{2}\,\psi_C,
\end{split}
\end{equation}
where $\psi_{C,X}$ are the complex order parameters of cavity photons ($C$) and quantum well excitons ($X$), $\epsilon^0_{C,X}$ the bottoms of their dispersions, $g>0$ the constant of exciton-exciton repulsive interaction. The imaginary terms in the right-hand sides read as follows: $\kappa$ is the photons decay rate, $\gamma=\gamma_X-\kappa_X$ is the effective linear gain rate of excitons, where $\gamma_X$ and $\kappa_X$ are the phonon-assisted exciton scattering from the reservoir and decay, respectively. $\Gamma|\psi_X|^2$ represents the exciton losses due to exciton-exciton scattering, effectively taking into account possible non-linear feeding of the condensate.
All energies are rescaled in the units of $\hbar\Omega_R$, lengths in the units of $\sqrt{\hbar/m_C\Omega_R}$, time in the units of $\Omega_R^{-1}$, and the wave functions in the units of $\sqrt{\hbar/m_C\Omega_R}$ ($\Omega_R$ is the Rabi coupling strength between the photon and exciton modes, and $m_C$ is the effective mass of cavity photon). The schematics of the model are presented in Fig.~\ref{fig_1}(a).

Using the Madelung form of the wave functions $\psi_{C,X}(t)=\sqrt{n_{C,X}(t)}\,e^{i\phi_{C,X}(t)}$ and introducing the new variables, $n(t)=n_C(t)+n_X(t)$, $\rho(t) = n_C(t)-n_X(t)$, and $\phi(t)=\phi_C(t)-\phi_X(t)$, one gets the set of motion equations
\begin{eqnarray}
\dot{n} &=& (\gamma-\kappa)n - (\gamma+\kappa)\rho-\frac{\Gamma}{2}(n - \rho)^2, \label{total} \\
\dot{\rho} &=& -\sqrt{n^2-\rho^2}\,\sin \phi - (\gamma+\kappa)n \nonumber\\
& & \qquad\qquad\qquad\qquad +(\gamma- \kappa)\rho+\frac{\Gamma}{2}(n-\rho)^2, \label{imbalance}
\\
\dot{\phi} &=&  -(\epsilon^0_C-\epsilon^0_X) + \frac{g(n-\rho)}{2} + \frac{\rho}{\sqrt{n^2-\rho^2}}\,\cos \phi.\label{phase}
\end{eqnarray}
The evolution equations (\ref{total})--(\ref{phase}) allows one to understand the effects of pumping and losses on the dynamics of the two-component condensate using the analogy of mechanical pendulum with tilt angle $\phi$ and gravity acting towards the lower polariton condesate state with $\phi=\pi$. The pendulum dynamics can then be visualized on a Bloch sphere of the radius $n$, which length is not conserved in the general case (see Fig.~\ref{fig_1}(b) and more explanations in Ref.\cite{SM}).

In the case of no pumping and losses ($\kappa,\gamma,\Gamma\equiv0$), Eq.(\ref{total}) immediately reduces to $\dot{n}=0$, while the equations (\ref{imbalance}), (\ref{phase}) acquire autonomous Hamiltonian form $\dot{\rho} = \partial H/\partial \phi$, $\dot{\phi} = -\partial H/\partial \rho$ with $H(\phi,\rho) = (\epsilon^0_C-\epsilon^0_X - gn/2)\rho + g\rho^2/4 + \sqrt{n^2-\rho^2}\,\cos \phi$. The condensate density $n$ is constant, and the influence of interactions on the dynamics of bounded motion in this case is negligible \cite{voronova_prl}. For zero detuning ($\epsilon^0_C=\epsilon^0_X$), the Hamilton equations can be expressed in terms of action-angle coordinates: $\dot{J}=0$, $\dot{\theta}=1$, with generalized conserved momenta
$J \equiv \frac{1}{2\pi}\oint\rho d\phi = n-|H|$.
The angular variable $\theta$ which is canonically conjugate to $J$ is introduced as polar angle in the phase space $(\rho,\phi)$, defining the position of the system on the orbit at a given moment of time. The pair of conjugate variables $\rho$ and $\phi$ can now be defined in terms of $(\theta,J)$:
\begin{equation}\label{rho_J}
\rho=\mp\sqrt{J(2n-J)}\sin\theta,
\end{equation}
\begin{equation}\label{S_J}
|\cos \phi|=\frac{n-J}{\sqrt{n^2-J(2n-J)\sin^2\theta}}.
\end{equation}

We use the expressions (\ref{rho_J}) and (\ref{S_J}) for approximate analytical investigation of the case when pump and dissipation are present. The transition from the variables $(n,\rho,\phi)$ to $(n,J,\theta)$ in the evolution equations (\ref{total})--(\ref{phase}) allows to separate fast and slow motion: since the phase variable $\theta$ changes fast while the variables $n$ and $J$ undergo slow evolution, it is justified to average the evolution equations for $\dot{n}(t)$ and $\dot{J}(t)$ over ``fast time'' (assuming ergodicity of the system, we replace time averaging with averaging over $\theta$) \cite{bogoliubov}, with the result
\begin{equation}\label{n_averaged}
\langle\dot{n}\rangle=(\gamma-\kappa)n-\frac{\Gamma}{2}\,n^2-\frac{\Gamma}{4}\,J(2n-J),
\end{equation}
\begin{equation}\label{J_averaged}
\langle\dot{J}\rangle=J\left[(\gamma-\kappa)-\Gamma\left(n-\frac{J}{4}\right)\right].
\end{equation}

\begin{figure}[t]
\renewcommand{\captionlabeldelim}{.}
\includegraphics[width=\columnwidth]{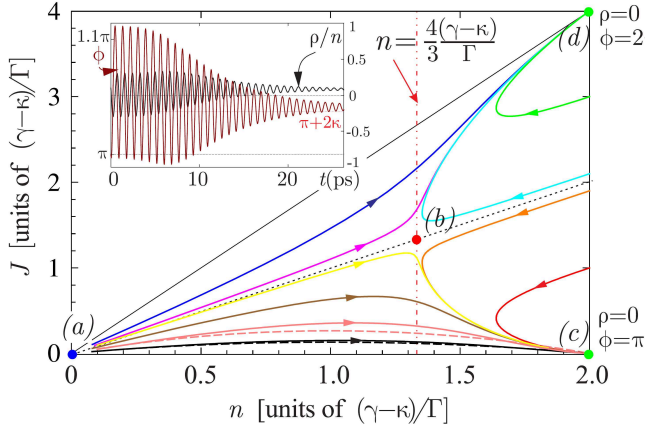}
\caption{\small (color online) Integral trajectories of the averaged set of equations (\ref{n_averaged}), (\ref{J_averaged}) on the phase plane $(n,J)$ (solid lines: numerical solutions; dashed lines: analytical solutions in the small amplitude limit). Fixed points: \textit{(a)} --- unstable node, \textit{(b)} --- saddle, \textit{(c)} and \textit{(d)} --- stable nodes (see the text for more details). Thin dotted line marks the separatrix $J=n$.
Vertical dot-dashed line $n=4(\gamma-\kappa)/3\Gamma$ shows the point for each trajectory when the population imbalance oscillations start to decay while the trajectories move away from the separatrix line. Inset shows numerical solutions $\rho(t)$ and $\phi(t)$ in the small amplitude limit, for $\kappa=0.1$, $\gamma=0.2=2\gamma^{\mbox{\footnotesize thr}}$, $g=0.002$, $\Gamma/g=1$. All energies are given in the units of $\hbar\Omega_R=5$~meV.}
\label{fig_Jn}
\end{figure}

Phase portrait of the averaged evolution equations (\ref{n_averaged}), (\ref{J_averaged}) is presented in Fig.~\ref{fig_Jn}. One can see that there are two sets of trajectories which are divided by the separatrix line $J=n$, with four fixed points:
unstable node $(0,0)$,
saddle $(4(\gamma-\kappa)/3\Gamma,4(\gamma-\kappa)/3\Gamma)$,
and two stable nodes, $(2(\gamma-\kappa)/\Gamma,0)$ and
$(2(\gamma-\kappa)/\Gamma,4(\gamma-\kappa)/\Gamma)$.
The system sets on a trajectory defined by the initial conditions $\rho(0)$ and $\phi(0)$, and gets attracted to one of the nodes, \textit{(c)} or \textit{(d)}, which correspond to equilibrium LP and UP condensates, respectively.
As follows from (\ref{rho_J}), $J=n$ corresponds to the maximal possible amplitude of population imbalance oscillations $\rho_m=n$, hence the closer the trajectory is to the separatrix, the larger is the amplitude of oscillations.  It is worth noting that after the averaging, all terms containing the interaction constant $g$ in the initial set of equations disappear. Therefore the Eqs. (\ref{n_averaged}), (\ref{J_averaged}) can be considered valid only for the cases when interactions are negligible.
To account for interactions, one has to consider higher approximation of the Krylov-Bogoliubov averaging method \cite{bogoliubov}.

In the case when amplitude of oscillations is small during the whole evolution time (which corresponds to integral trajectories far from the separatrix in Fig.~\ref{fig_Jn}), the set of equations (\ref{total})--(\ref{phase}) allows analytical solution. Imposing $\rho\ll n$ in (\ref{total}) or, equivalently, $J\ll n$ in (\ref{n_averaged}), one gets
$n(t)=(\gamma-\kappa)/\Gamma\left[1+\tanh\{(\gamma-\kappa)(t+t_0)/2\}\right]$,
where $t_0 = 2\arctanh(\Gamma n(0)/(\gamma-\kappa)-1)/(\gamma-\kappa)$. At $t\rightarrow\infty$, this solution gives the limiting value for the condensate population, $n_\infty=2(\gamma-\kappa)/\Gamma$.
Similarly, assuming $J\ll n$ in (\ref{J_averaged}), one gets
$J = J(0)\cosh^2((\gamma-\kappa)t_0/2)/\cosh^2((\gamma-\kappa)(t+t_0)/2)$.
The analytical solutions for trajectories corresponding to the small amplitude limit are plotted in Fig.~\ref{fig_Jn} as dashed lines.

Using the stable node coordinates $\rho=0$, $\phi=\pi$ (see Fig.~\ref{fig_Jn}) as a starting point of unaveraged evolution analysis, we linearize the Eqs.~(\ref{imbalance}) and (\ref{phase}) in the region $\rho/n\ll1$ and $|\phi-\pi|\ll1$. In adiabatic approximation, assuming $n(t)$ a known, slowly changing function, we get damped-driven pendulum equations for $\rho(t)$ and $\phi(t)$. From those equations we extract the damping rates of population imbalance $\beta_\rho=3(\Gamma n)/4-(\gamma-\kappa)$ and relative phase $\beta_\phi=(\gamma-\kappa-\Gamma n/2)/(2+gn)$
and the new approximate coordinates for the fixed point (focus):
\begin{equation}\label{inf_values}
\left(\frac{\rho}{n}\right)_\infty\simeq\frac{1}{1+\frac{\Gamma}{g(\gamma-\kappa)}}\,; \quad \phi_\infty\simeq\pi+2\kappa.
 \end{equation}

\begin{figure}[b]
\renewcommand{\captionlabeldelim}{.}
\includegraphics[width=\columnwidth]{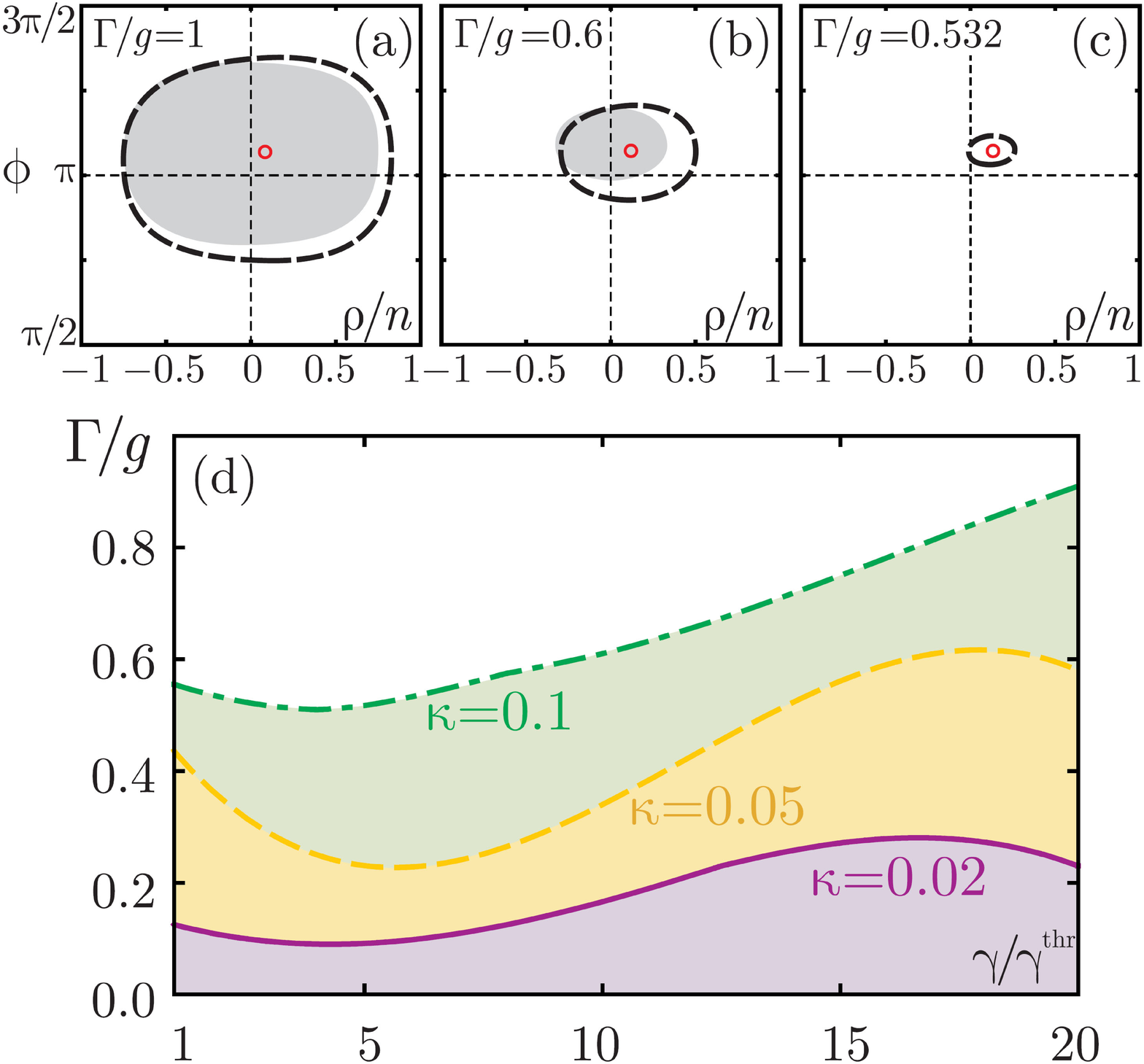}
\caption{\small (color online) (a)--(c) Basin of attraction of the lower fixed point (red circle) and projection of the saddle limit cycle (black dashed line) for $\Gamma/g$ as indicated on the panels and other parameters same as in Fig.~\ref{fig_Jn}. Initial values of $(\rho,\phi)$ that get attracted to the fixed point (\ref{inf_values}) are colored grey, others which flow towards the limit cycle and get attracted to the upper fixed point (see Fig.~\ref{difG}(a)) are white. (d) Bifurcation diagram showing the regions of stability of the lower fixed point and existence of the trajectories flowing towards the LP condensate, for three values of decay rate $\kappa$ as marked. Regions above the lines correspond to the existence of points flowing towards the LP condensate. For colored regions below the lines, any starting point gets attracted to the upper focus.}
\label{bif}
\end{figure}

Numerical solutions for $\rho(t)$ and $\phi(t)$ in the small amplitude limit are shown in the inset of Fig.~\ref{fig_Jn}. The analytical description we developed reveals amplitude-dependent damping, bringing the analogy with Van der Pol oscillator. The oscillations are amplified as long as $n<4(\gamma-\kappa)/3\Gamma$,
and in the integral phase portrait $(n,J)$ each trajectory is approaching the saddle point at the separatrix. On the phase plane $(\rho,\phi)$ it corresponds to evolution lines flowing towards the limit cycle, where damping $\beta_\rho$ goes to zero. In 3D phase space $(n,\rho,\phi)$ this is a saddle limit cycle \cite{SM}. Its projection on the plane $(\rho,\phi)$ is shown in Fig.~\ref{bif}(a)--(c): after approaching the limit cycle, all trajectories flow away, and oscillations start to relax towards one of the two stable fixed points shown in Fig.~\ref{fig_Jn}.

\begin{figure}[b]
\renewcommand{\captionlabeldelim}{.}
\includegraphics[width=\columnwidth]{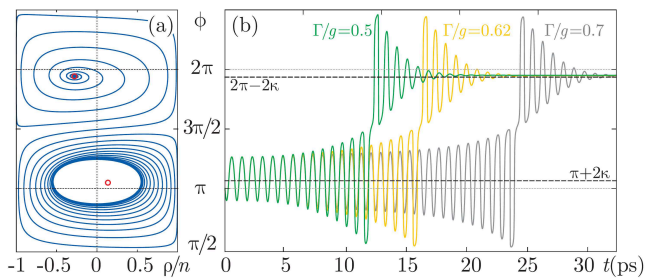}
\caption{\small (color online) (a) Phase-plane projection of a trajectory starting near the LP state, which is getting attracted to the UP state, for $\Gamma/g=0.7$, $\rho(0)=0.5n$, $\phi(0)=\pi$. Other parameters are the same as in Fig.~\ref{fig_Jn}. Red circles mark the two stable foci corresponding to the LP and UP equilibrium states. (b) Relative phase $\phi$ against time for three values of $\Gamma/g$ as marked.}
\label{difG}
\end{figure}

Given $2\kappa\ll1$ and $g(\gamma-\kappa)/\Gamma\ll1$, the averaged equations (\ref{n_averaged}), (\ref{J_averaged}) give results which are in a good agreement with the solutions of the full system (\ref{total})--(\ref{phase}). However, with the increase of pumping and decay rates, one can only use general propositions of the theory of dynamical systems, and numerical analysis. For arbitrary values of parameters we linearize the equations (\ref{total})--(\ref{phase}) around the fixed points and study the properties of the obtained linear operators. Far from the fixed points, we model the evolution numerically. We find that while the upper focus (point \textit{(d)} in Fig.~\ref{fig_Jn}) stays always stable, for the lower focus (point \textit{(c)}) each pair $(\kappa,\gamma)$ reveals values of $\Gamma$ at which the pair of complex conjugate eigenvalues of the linearization crosses the imaginary axis. This behavior is known as Hopf bifurcation and it corresponds to disappearance of the limit cycle while the lower fixed point loses stability.
As shown in Fig.~\ref{bif}(a)--(c), for large values of $\Gamma$ (in the units of $g$), there can be two regimes of the dynamics depending on the starting point of the phase-plane ($\rho,\phi)$: if the trajectory starts from the basin of attraction of the lower fixed point, the dynamics is that of relaxation oscillations shown in the inset of  Fig.~\ref{fig_Jn}. The system then relaxes towards the equilibrium LP condensate ($\phi\simeq\pi$). Contrary to this, if the system is prepared in the initial state lying outside the basin of attraction, the trajectory will flow towards the limit cycle, and then through the series of large-amplitude oscillations wind up towards the upper focus, which corresponds to the upper polariton condensate ($\phi\simeq2\pi k$, $k\in\mathds{Z}$). The trajectory projection on the phase plane for the latter case is shown in Fig.~\ref{difG}(a). With the decrease of $\Gamma$, the basin of attraction shrinks, and at some critical value of $\Gamma$, the LP state becomes unstable, the limit cycle disappears, and all trajectories get attracted to the UP state (see Fig.~\ref{difG}(a)). The bifurcation diagram plotted in Fig.~\ref{bif}(d) for three values of $\kappa$ shows the parameters at which the system will stabilize in the UP state regardless of the initial conditions. In the pendulum analogy, transition to the upper equilibrium can be compared with Kapitza pendulum  which stabilizes in the state upwards while gravity is acting downwards \cite{kapitza}. However, for Kapitza pendulum, the inverted state stabilizes due to fast vibrations of the suspension point, while in our case the amplification towards the limit cycle and consequent stabilization of the ``inverted pendulum'' state happen due to oscillations of the pendulum length (the radius of the Bloch sphere).

The formation of the ``inverted pendulum'' state of a polariton Rabi oscillator can be physically understood as follows.
Due to the blueshift of the exciton mode with respect to the photon mode induced by the exciton pumping, the upper polariton state becomes more ``exciton-like'', while the lower polariton state becomes more ``photon-like''. Consequently, the excitonic pumping mostly feeds the upper state, while the lower state is emptied due to the photonic leakage through the Bragg mirrors. As a result, the exciton population increases and the photon population decreases. This process is limited and stabilized by the non-linear exciton dissipation process. The dynamics is then determined by the competition of real, $g|\psi_X|^2$, and imaginary, $\Gamma|\psi_X|^2$, non-linearities in Eq.~(\ref{cGL_1}). When $\Gamma$ is large enough compared to $g$, the system relaxes towards the LP condensate. When $\Gamma$ is low, however, the excitonic fraction of the polariton gas grows, which leads to the accumulation of particles on the UP branch.
The times at which this stabilization occurs for given initial conditions also depend on the ratio $\Gamma/g$ (see Fig.~\ref{difG}(b)). Note that while transiting between the two equilibria, the system passes though the so-called ``internal Josephson'' regime of the running relative phase \cite{voronova_prl,SM}.

In conclusion, we have demonstrated theoretically the existence of an ``unstable Val der Pol'' limit cycle and a new inverted stationary state of a polariton Rabi oscillator similar to classical Kapitza pendulum. Realization of this state requires excitonic pumping that would exceed losses due to the exciton scattering with acoustic phonons, and relatively small non-linear losses due to exciton-exciton scattering. The described effects are essentially non-linear: the stationary populations of upper and lower polariton states are dependent on the balance between exciton pumping and non-linear losses.
It is worth mentioning that the Kapitza pendulum effect was also shown for atomic condensates in oscillating double-well potentials \cite{boukobza}, however, here it has a completely different nature.

We would like to thank A.~Kavokin for discussions. The work of NSV is financially supported by Russian Foundation for Basic Research (RFBR research project No. 16--32--60066 mol\_а\_dk). YuEL is supported by Program of Basic Research of High School of Economy.

\newpage
\begin{widetext}
\center{
\large{\bf Inverted pendulum state of a polariton Rabi oscillator \\[10pt] Supplemental Material}
}
\vspace{10pt}
\begin{quote}
In this supplementary material, we provide details on the parameters used in the simulations, visualization of the internal dynamics on the Bloch sphere, extended analytical analysis of the limiting case and stability, and additional numerical data.
\end{quote}
\vspace{10pt}
\end{widetext}

\section{I. On the pendulum analogy}

The physics of a Bosonic Josephson junction \cite{bjj,sols,oberthaler} is described by the non-linear autonomous equations of motion for population imbalance $\rho(t) = n_1-n_2$ and relative phase $\phi(t)=\phi_1-\phi_2$ between the 1st and 2nd condensates, which are characteristic for a nonrigid pendulum:
\begin{equation}\label{autonomous}
\begin{split}
\dot{\rho} &= -\sqrt{n^2-\rho^2}\,\sin \phi, \\
\dot{\phi} &=  -\Delta E -\Lambda\,\rho + \frac{\rho}{\sqrt{n^2-\rho^2}}\,\cos \phi.
\end{split}
\end{equation}
With total amount of particles $n=n_1+n_2$ being conserved, the parameters $\Delta E$ and $\Lambda$ accounting for asymmetry of the trap and interactions, respectively, determine the dynamics of the system. The nonlinear effects such as macroscopic quantum self-trapping\cite{smerzi} and $\pi$-phase modes  \cite{raghavan} which are not observable with the superconducting junctions arise from (i) interactions and (ii) nonrigidity (momentum-dependent length) of the pendulum.

In our model, the motion equations for $\rho$ and $\phi$ (4) and (5) of the main text, compared to the equations (\ref{autonomous}) with $\Delta E=\epsilon^0_C-\epsilon^0_X - gn/2$ and $\Lambda = gn/2$, contain dissipation as well as external driving field, since the total population $n(t)$ is changing.

\begin{figure}[t]
\renewcommand{\captionlabeldelim}{.}
\includegraphics[width=\columnwidth]{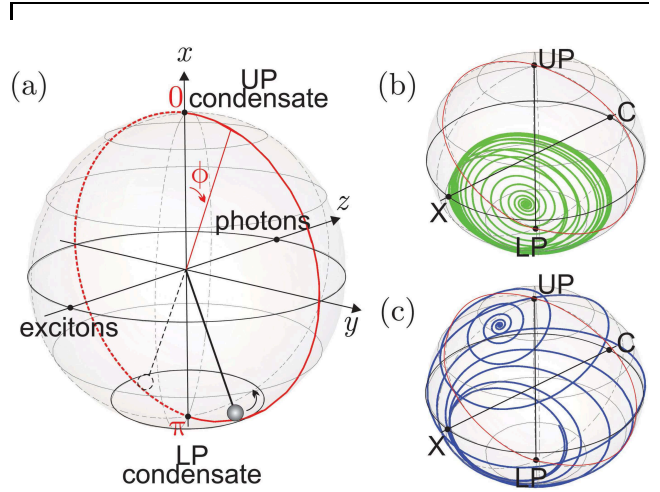}
\caption{\small (a) Bloch sphere of the radius $n$ rotated by $\pi/2$ and the pendulum analogy for the problem. Azimuthal angle on the sphere (tilt angle of the pendulum) corresponds to relative phase $\phi$ changing from 0 to $2\pi$, polar angle $\theta$ changing from 0 to $\pi$ corresponds to normalized population imbalance: $\cos\theta=\rho/n$. (b)--(c) Evolution trajectories on the normalized Bloch sphere, both starting in the point $\rho(0)=0.5n$, $\phi(0)=\pi$ for $\kappa=0.1$, $\gamma=0.4=4\gamma^{\mbox{\footnotesize thr}}$, $g=0.002$. (b) $\Gamma/g=1$; the system relaxes towards the LP condensate, to the point given by (\ref{focus_LP}). (c) $\Gamma/g=0.8$; the stabilization of ``inverted pendulum'' state on the UP branch, the focus point is given by (\ref{focus_UP}). All energies are in the units of $\hbar\Omega_R=5$~meV.}
\label{fig_spheres}
\end{figure}

For better understanding of the effects of pumping and losses on the dynamics of the two-component condensate, we visualize total population $n$ as a Bloch sphere radius with coordinates $x=\sqrt{n^2-\rho^2}\cos \phi$, $y=\sqrt{n^2-\rho^2}\sin \phi$, $z=\rho$, which length $\sqrt{x^2+y^2+z^2}$ is not conserved in the general case. To merge this picture with the pendulum analogy, one would have to rotate the sphere by $\pi/2$ so that the poles corresponding to the pure photonic (north) and pure excitonic (south) states are located at the new equator. Then the relative phase $\phi$ rotating along the old equator coincides with the tilt angle of the pendulum. Note that due to the positive sign chosen in front of the coupling terms in Eqs. (1) of the main text, the gravity of the pendulum acts towards the state $\phi=\pi$ corresponding to the condensate of lower polaritons, and the state $\phi=0$ is at the new north pole corresponding to the upper polariton condensate (see Fig.~\ref{fig_spheres}(a)). Population imbalance $\rho$ is connected to the polar angle on the sphere as $\rho/n=\cos\theta$. This representation matches with the one used in Ref.~\cite{ultrafast}.

\section{II. Parameters}

We assume that at the moment $t=0$ the two-component condensate is formed by a short pulse of light exciting both LP and UP branches, setting the initial populations and phases, and will impose for all our simulations $n(0)=1$ (which in scaled units corresponds to $\sim10^{10}$~cm$^{-2}$) and $(\rho(0),\phi(0))$ on the lower semi-sphere of Fig.~\ref{fig_spheres}(a). Initial value of the action variable $J$ can be defined from the Eqs.~(5) and (6) of the main text, and reads $J(0)=n(0)-\sqrt{n^2(0)-\rho^2(0)}\cos S(0)$. Note that for the sake of simplicity, we consider only the case of zero energy detuning $\epsilon^0_C-\epsilon^0_X=0$.

Before describing the different regimes of dynamics, it is convenient to approximately define the parameters values.
We take the interaction constant $g=0.002$ (in scaled units, $g_X=0.015$~meV$\cdot\mu$m$^2$). Values of the decay rate $\kappa$ in our simulations vary from 0.02 to 0.1 (which corresponds to $\kappa_C$ from $0.1$ to $0.5$~meV). This is consistent with the experiments to date reporting the cavity photon linewidth at low excitation powers, see e.g. Ref. \cite{kasprzak}. The linear gain rate for excitons at threshold is equal to $\gamma^{\mbox{\footnotesize thr}}=\kappa$, so we use the values from just above threshold up to 20 times the threshold pump power, $1<\gamma/\gamma^{\mbox{\footnotesize thr}}<20$. Obviously, if $\gamma<\kappa$, the condensate density will decay to zero. At last, the value of the saturation coefficient $\Gamma$, which is most important as it brings the imaginary nonlinearity to the system, is unclear. If $\Gamma$ is too small, the condensate population $n$ grows (given $\gamma>\kappa$) until reaching the equilibrium value $n_\infty=\gamma/\Gamma$ and the population imbalance $\rho$ shifts to $-n$ as all photons leak out of the cavity and the system becomes filled with excitons. If $\Gamma$ is very high (larger than the interaction constant $g$), \textit{i.e.} the saturation is fast, the total population tends to the equilibrium value $n_\infty=2(\gamma-\kappa)/\Gamma$ while the (normalized) population imbalance performs small-amplitude, fast decaying oscillations around $gn_\infty/2$. At further increase of $\Gamma$, the dynamics does not change except for speeding-up of the oscillations decay. As we indeed see in our simulations, the ratio $\Gamma/g$ is one of the values determining the type of dynamics. It has been argued \cite{berloff,eastham} that the imaginary nonlinearity $\Gamma |\psi_X|^2$ in Eq.~(1) of the main text should be 3 to 40 times smaller than the real nonlinearity $g|\psi_X|^2$. For the present discussion, we will use the values in the range $0.025<\Gamma/g<1$.

\section{III. Analytical investigation}\label{analyt}

To get approximate analytical description of the dynamics, we consider the limiting case of small-amplitude oscillations. Assuming $\rho\ll n$ in (\ref{surface}) gives $n_\infty=2(\gamma-\kappa)/\Gamma$, while the linearization of the Eqs. (3) and (4) of the main text around the approximate equilibrium value $\rho=0$, $\phi=\pi$ (the lower fixed point) found in the analysis of the averaged evolution leads to the equations of damped-driven oscillations as follows:
\begin{widetext}
\begin{equation}\label{rho_lin}
\ddot{\rho}+2\left[\frac{3}{4}\,\Gamma n - (\gamma-\kappa)\right]\dot{\rho} + \left[1+\frac{gn}{2}+\Gamma\kappa n + (\gamma-\kappa)^2\right]\rho=n\left[\frac{gn}{2}+\frac{\Gamma n}{2}\,(\gamma-\kappa)-\left(\frac{\Gamma n}{2}\right)^2\right],
\end{equation}
\begin{equation}\label{phi_lin}
\ddot{\phi}+2\,\frac{\gamma-\kappa-\Gamma n/2}{2+gn}\,\dot{\phi}+\left(1+ \frac{gn}{2}\right)\phi = \left[\pi + \gamma + \kappa -\frac{\Gamma n}{2}\right]\left(1+ \frac{gn}{2}\right)-\frac{gn+(\frac{gn}{2})^2}{1+ \frac{gn}{2}} \left[\gamma-\kappa-\frac{\Gamma n}{2}\right].
\end{equation}
\end{widetext}
All the coefficients of Eqs. (\ref{rho_lin}), (\ref{phi_lin}) are time-dependent, with $n(t)$ changing monotonically according to the law derived in the main text:
\begin{equation}\label{n_analyt}
n(t)=\frac{\gamma-\kappa}{\Gamma}\left[1+\tanh\left(\frac{\gamma-\kappa}{2}(t+t_0)\right)\right]
\end{equation}
with $t_0 = 2\arctanh(\Gamma n(0)/(\gamma-\kappa)-1)/(\gamma-\kappa)$.

It is worth noting that while the natural frequency of oscillations of the relative phase $\phi$ stays unchanged as compared to the conservative case of no gain and dissipation (see Ref.\cite{voronova_prl}), frequency of population imbalance acquires additional corrections due to non-zero $\gamma$, $\kappa$ and $\Gamma$. More, the damping of the oscillations is density-dependent: as immediately follows from (\ref{rho_lin}), as long as $n<4(\gamma-\kappa)/3\Gamma$ the oscillations of population imbalance are amplified, and only after $n$ passes the saddle point shown in Fig.~2 of the main text they start to decay.
Oscillations of the relative phase are damped for all $n<n_\infty=2(\gamma-\kappa)/\Gamma$.

Noticing that for $\rho\ll n$ we obtained $n_\infty\simeq2(\gamma-\kappa)/\Gamma$, one finally finds the approximate coordinates of the lower focus:
\begin{equation}\label{focus_LP}
\left(\frac{\rho}{n}\right)_\infty^{\mbox{\tiny LP}}\simeq\frac{1}{1+\frac{\Gamma}{g(\gamma-\kappa)}}
\,, \quad\phi_\infty^{\mbox{\tiny LP}}\simeq\pi+2\kappa.
 \end{equation}
Similarly linearizing the evolution equations around the upper fixed point of the averaged dynamics, $\rho=0$, $\phi=2\pi k$ ($k\in\mathds{Z}$), for the upper focus one has
\begin{equation}\label{focus_UP}
\left(\frac{\rho}{n}\right)_\infty^{\mbox{\tiny UP}}\simeq-\frac{1}{1+\frac{\Gamma}{g(\gamma-\kappa)}}
\,, \quad\phi_\infty^{\mbox{\tiny UP}}\simeq2\pi k-2\kappa.
\end{equation}

As follows from these results, the regime of small-amplitude damped oscillations can take place only if $2\kappa\ll1$ and $g(\gamma-\kappa)/\Gamma\ll1$, \textit{i.e.} at low decay rate, low pump powers and relatively large saturation coefficients (comparable to $g$). Given $\kappa$ is small, if $\Gamma/g$ is decreased at a fixed value of $\gamma$ or, alternatively, the gain rate $\gamma$ is increased at fixed $\Gamma/g$, the initial assumption $\rho/n\ll 1$ and, subsequently, the equations (\ref{rho_lin}) and (\ref{phi_lin}) will be invalid. Note, however, that for the case of no interactions ($g=0$), for arbitrary values of the parameters one has  $n\rightarrow 2(\gamma-\kappa)/\Gamma$, $\langle\rho\rangle=0$ and $\langle S\rangle=\pi+2\kappa$, even for large $\rho(0)$ comparable to $n(0)$ (then the initial amplitude of oscillations is not small).

\begin{figure}[b]
\renewcommand{\captionlabeldelim}{.}
\includegraphics[width=\columnwidth]{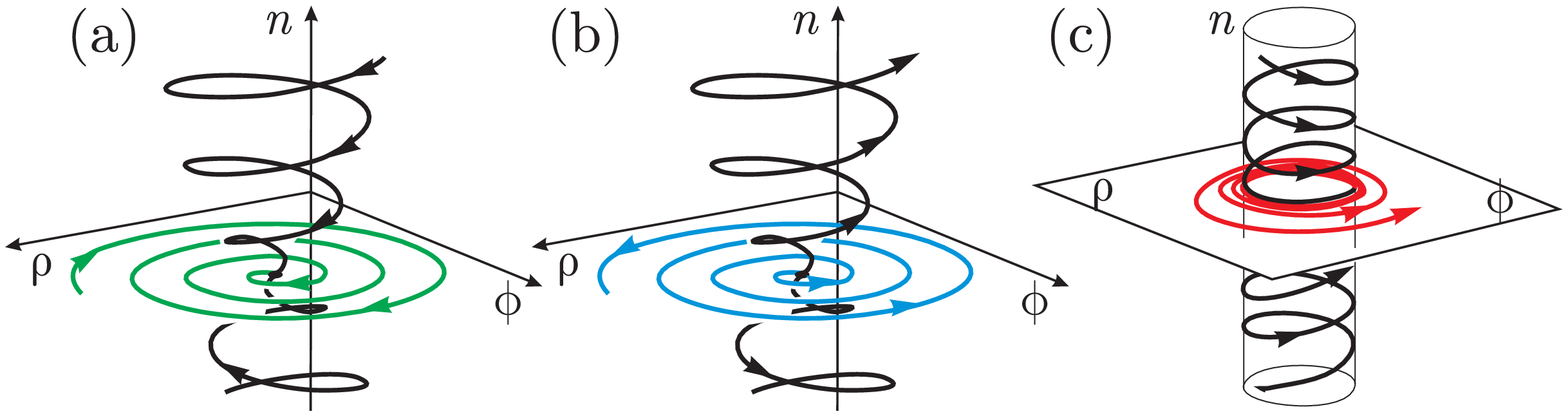}
\caption{\small Illustration of the trajectories (black lines) in 3D phase space $(n,\rho,\phi)$ and their projections (colored lines) on the phase plane $(\rho,\phi)$ when approaching (a) a stable focus; (b) an unstable focus; (c) a saddle limit cycle.}
\label{equilibria}
\end{figure}

To obtain information about the stability of the fixed points, we graphically determine the exact foci coordinates $n_\infty$, $\rho_\infty$, and $\phi_\infty$ by imposing $\dot{n}=0$, $\dot{\rho}=0$, and $\dot{\phi}=0$ for a steady state at $t\rightarrow\infty$ in the evolution equations (2)--(4) of the main text. Each possible evolution line ends at a point on the resulting surface $\rho(n)$ in the 3D-space $(n,\rho,\phi)$
\begin{equation}\label{surface}
(\gamma-\kappa)n-(\gamma+\kappa)\rho-\frac{\Gamma}{2}(n-\rho)^2=0.
\end{equation}
Intersection of this surface with the lines
\begin{multline}
\pm\sqrt{n^2-\rho^2}\sqrt{1-\frac{g^2}{4}\frac{(n-\rho)^2(n^2-\rho^2)}{\rho^2}}\\
+(\gamma+\kappa)n- (\gamma-\kappa)\rho+\frac{\Gamma}{2}(n-\rho)^2=0
\end{multline}
gives the coordinates $(n,\rho,\phi)$ of the two equilibria of the dynamical system. Linearizing the system in the vicinity of the fixed points, one gets a cubic equation for the three eigenvalues of the Jacobian matrix:
\begin{widetext}
\begin{multline}
\lambda^3 - 2(\gamma - \kappa - \Gamma n + \Gamma \rho) \lambda^2 - \left[(\gamma + \kappa - \Gamma n + \Gamma \rho) \frac{n}{\sqrt{n^2 - \rho^2}}\sin{\phi} \right.\\
\left.+(\gamma - \kappa - \Gamma n + \Gamma \rho) \frac{\rho}{\sqrt{n^2 - \rho^2}} \sin{\phi} + \frac{g}{2} \sqrt{n^2 - \rho^2} \cos{\phi} - \frac{n^2 \cos^2{\phi} - \rho^2 \sin^2{\phi}}{n^2 - \rho^2} + 4 \kappa (\gamma -  \Gamma n + \Gamma \rho)\right] \lambda \\
-  4\kappa (\gamma - \Gamma n + \Gamma \rho) \frac{\rho}{\sqrt{n^2 - \rho^2}} \sin{\phi} - \kappa g \sqrt{n^2 - \rho^2}\cos{\phi} - \frac{(\gamma - \kappa)n - (\gamma + \kappa) \rho - \Gamma(n - \rho)^2}{n^2 - \rho^2}\, n \cos^2{\phi}\\
 - \frac{(\gamma + \kappa)n - (\gamma - \kappa) \rho - \Gamma(n - \rho)^2}{n^2 - \rho^2}\,\rho \sin^2{\phi} = 0,
\end{multline}
\end{widetext}
which has one real root $\lambda_1<0$ and two complex conjugate roots $\lambda_{2,3}$. If $\lambda_1<0$ and $\mbox{Re}(\lambda_{2,3})<0$, then the focus in 3D-space is stable (attracting the trajectories), and if $\mbox{Re}(\lambda_{2,3})>0$, then the focus is unstable (repulsing the trajectories). The illustration is given in Fig.~\ref{equilibria}(a)--(b).
Changing the parameters and defining the coordinates ($n_\infty$, $\rho_\infty$, $\phi_\infty$) of the fixed points, we look for the values of $\gamma$, $\kappa$, and $\Gamma$ at which the complex eigenvalues cross the imaginary axis (i.e. $\mbox{Re}(\lambda_{2,3})=0$). For all values of the parameters that we consider, we find that the upper focus given approximately by (\ref{focus_UP}) is always stable, while the lower focus (\ref{focus_LP}) changes its stability.

As mentioned above, as long as density is small, damping $\beta_\rho$ approximately defined from (\ref{phi_lin}) is negative, and it reaches zero when $n(t)$ passes the saddle point on the averaged diagram (see Fig.~2 of the main text). For unaveraged trajectories in 3D phase space $(n,\rho,\phi)$ this point corresponds to a \textit{saddle limit cycle} (schematically shown in Fig.~\ref{equilibria}(c)). When trajectories approach the saddle limit cycle, $\langle n\rangle$ stays approximately constant, while on the phase plane $(\rho,\phi)$ the system is orbiting the same line without damping, then gets repulsed from it to get finally attracted to one of the stable equilibria (see the simulation results in Fig.~\ref{SFig_1}). The limit cycle exists as long as there are points being attracted to the lower equilibrium, and disappears in the moment when it loses stability. For trajectories projections on the 2D phase plane $(\rho,\phi)$, one then has a subcritical Hopf bifurcation in which a small-amplitude limit cycle is branching from a fixed point which changes type of stability (for more details see e.g. Ref.\cite{sagdeev}). The bifurcation diagram for this behavior is given in Fig.~3 of the main text.

\section{IV. Numerical results}
\subsection{Small-amplitude oscillations}

\begin{figure}[b]
\renewcommand{\captionlabeldelim}{.}
\includegraphics[width=\columnwidth]{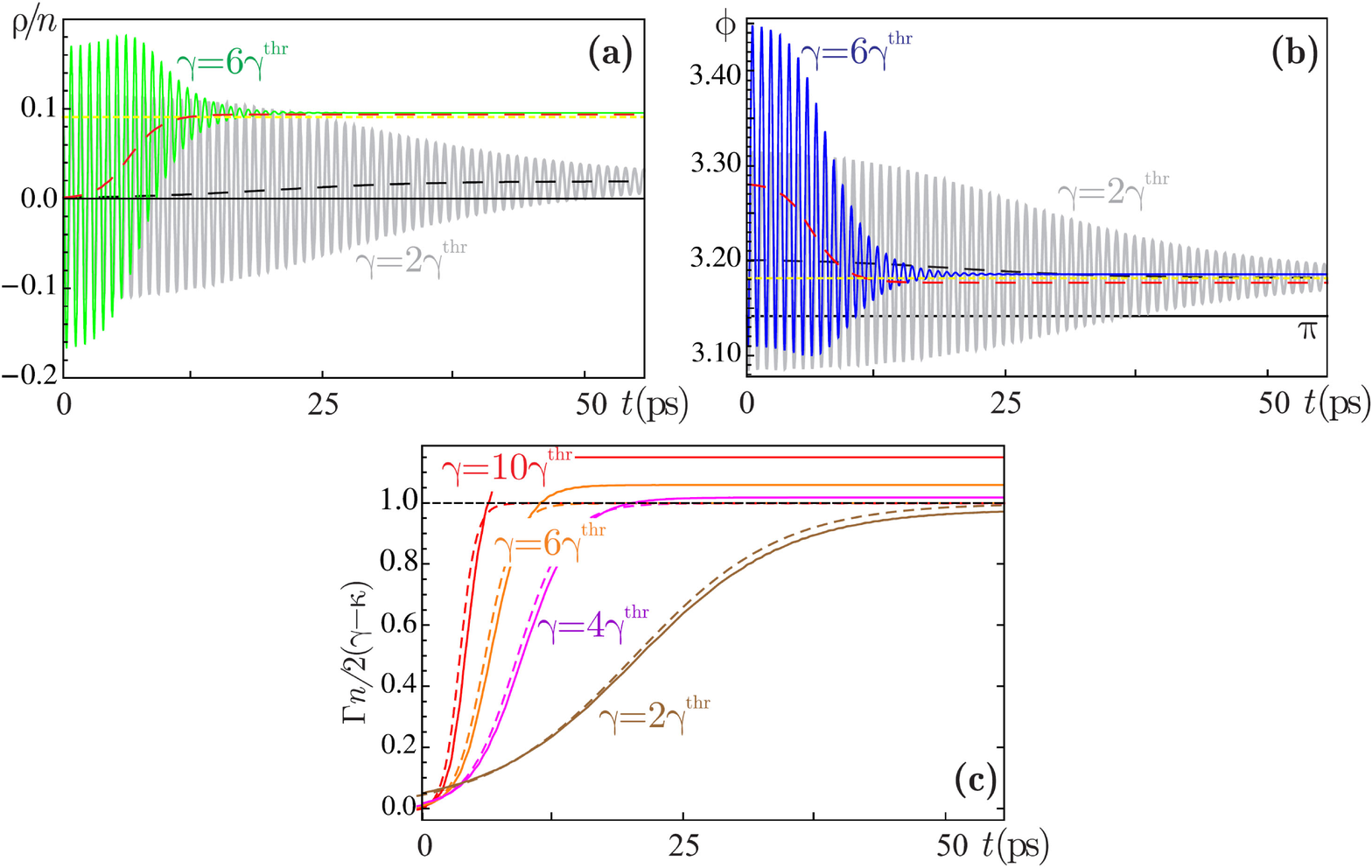}
\caption{\small (a) Normalized population imbalance $\rho/n$ and (b) relative phase $\phi$ as functions of time, for gain rates $\gamma=2\gamma^{\mbox{\footnotesize thr}}$ and $6\gamma^{\mbox{\footnotesize thr}}$ as marked. Dashed lines correspond to the density-dependent average values given by the right-hand sides of the Eqs. (\ref{rho_lin}) and (\ref{phi_lin}) for each $\gamma$. Thin light dotted lines mark the approximate average values $(\rho/n)_\infty^{\mbox{\tiny LP}}$ and $\phi_\infty^{\mbox{\tiny LP}}$ given by (\ref{focus_LP}). (c) Total polariton population $n$ against time for $\gamma=2\gamma^{\mbox{\footnotesize thr}}$, $4\gamma^{\mbox{\footnotesize thr}}$, $6\gamma^{\mbox{\footnotesize thr}}$ and 10$\gamma^{\mbox{\footnotesize thr}}$, plotted in the units of $2(\gamma-\kappa)/\Gamma$. Black dotted line at the level of unity corresponds to $n_\infty=2(\gamma-\kappa)/\Gamma$. Values of physical parameters used: $\hbar\Omega_R=5$~meV, $g_X=0.015$~meV$\cdot\mu$m$^2$, $\kappa_C=0.1$~meV, $\Gamma/g=1$.}
\label{small-gamma}
\end{figure}

We solve the set of equations (2)--(4) of the main text numerically for a wide range of the parameters in consideration. The simulations results agree with our analytical predictions for $\kappa$ and $g(\gamma-\kappa)/\Gamma$ of the order of $10^{-2}$. As the values of the parameters increase, the equilibrium value $n_\infty$ shifts away from $2(\gamma-\kappa)/\Gamma$ while $\rho(t)$ and $\phi(t)$ oscillate around the non-constant average values given by the right-hand sides of the Eqs. (\ref{rho_lin}), (\ref{phi_lin}).

The change of dynamical behavior brought by the increase of pump power is displayed in Fig.~\ref{small-gamma}. The panels (a) and (b) show the comparison of population imbalance and relative phase oscillations for $\gamma=2\gamma^{\mbox{\footnotesize thr}}$ and $6\gamma^{\mbox{\footnotesize thr}}$ at fixed $\Gamma/g=1$ and the dimensionless decay rate $\kappa=0.02$. Fig.~\ref{small-gamma}(c) shows total population $n(t)$ in the units of $2(\gamma-\kappa)/\Gamma$ for different values of $\gamma$. As can be seen, at the increase of $\gamma$, the limiting value $n_\infty$ shifts up from the value $2(\gamma-\kappa)/\Gamma$, while the values $\rho_\infty^{\mbox{\tiny LP}}$ and $\phi_\infty^{\mbox{\tiny LP}}$ are still approximately given by (\ref{focus_LP}). The lifetime of damped oscillations noticeably shortens with the increase of $\gamma$.

\begin{figure}[b]
\renewcommand{\captionlabeldelim}{.}
\includegraphics[width=\columnwidth]{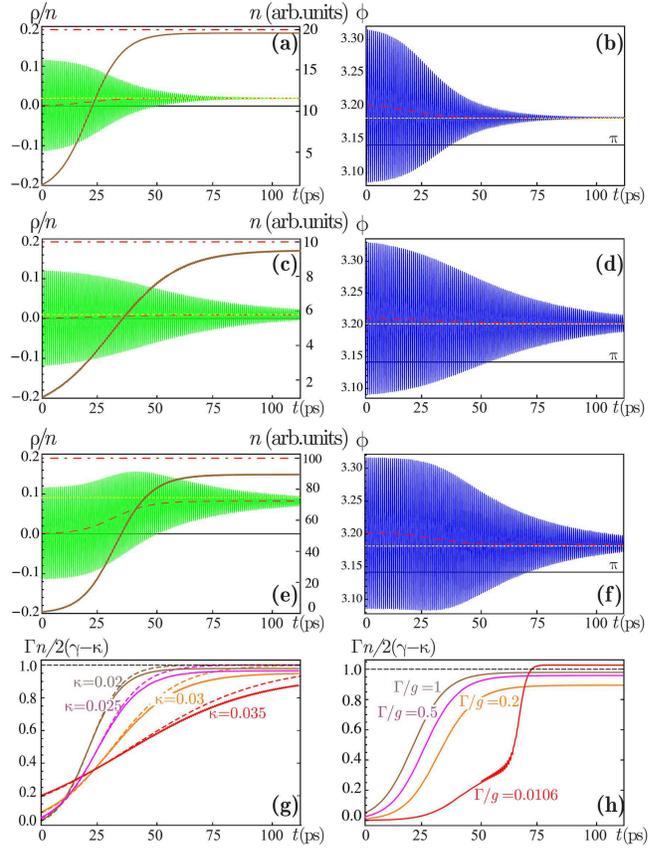}
\caption{\small (a)--(f) Normalized population imbalance $\rho/n$, total number of particles $n$ and relative phase $\phi$ as functions of time. For all panels $\gamma=0.04$ (which corresponds to $0.2$~meV). Dashed lines correspond to the time-dependent average values given by the right-hand sides of the Eqs. (\ref{rho_lin}) and (\ref{phi_lin}), thin light dotted lines correspond to $(\rho/n)_\infty^{\mbox{\tiny LP}}$ and $\phi_\infty^{\mbox{\tiny LP}}$ given by (\ref{focus_LP}). Dot-dashed line shows the limiting $n_\infty=2(\gamma-\kappa)/\Gamma$. (a) and (b) $\kappa=0.02$, $\Gamma/g=1$; (c) and (d) $\kappa=0.03$, $\Gamma/g=1$; (e) and (f) $\kappa=0.02$, $\Gamma/g=0.2$. See the text for more explanations.
(g) Total population $n$ against time plotted in units of $2(\gamma-\kappa)/\Gamma$ for decay rates $\kappa=0.02$, $0.025$, $0.03$, $0.035$ as marked at fixed $\gamma=0.04$ and $\Gamma/g=1$. Black dotted line at the level of unity corresponds to $n_\infty=2(\gamma-\kappa)/\Gamma$. (h) Same for fixed values of $\kappa=0.02$ and $\gamma=0.04=2\gamma^{\mbox{\footnotesize thr}}$, $\Gamma/g=1$, $0.5$, $0.2$ and $0.106415$ as marked. Other parameters are the same as used in Fig.~\ref{small-gamma}.}
\label{small-ampl}
\end{figure}

Fig.~\ref{small-ampl} shows the simulations results for different values of the parameters $\kappa$ and $\Gamma$ at a fixed value of gain rate $\gamma$. The figure should be analyzed as follows. The upper panels (a) and (b) show the same case as grey curves in Fig.~\ref{small-gamma}(a),(b), which are in perfect agreement with the analytical analysis made for the case when $2\kappa\ll1$ and $g(\gamma-\kappa)/\Gamma\ll1$. The next two pairs of panels should be separately compared to the upper pair. In Fig.~\ref{small-ampl}(c),(d), the decay rate $\kappa$ is increased at a fixed $\Gamma/g$. In Fig.~\ref{small-ampl}(e),(h), $\kappa$ is kept the same as in (a),(b) while the ratio $\Gamma/g$ is taken 5 times smaller. For both cases, the lifetime of oscillations increases while $n_\infty$ becomes smaller than $2(\gamma-\kappa)/\Gamma$ and the time-average values $(\rho/n)_\infty^{\mbox{\tiny LP}}$, $\phi_\infty^{\mbox{\tiny LP}}$ shift from those given by (\ref{focus_LP}). Note that for all cases, the damped Rabi oscillations follow the time-dependent average lines given by the right-hand sides of the Eqs. (\ref{rho_lin}), (\ref{phi_lin}). Fig.~\ref{small-ampl}(g) and (h) show the total population $n(t)$ in the units of $2(\gamma-\kappa)/\Gamma$ for several decay rates $\kappa$ and ratios $\Gamma/g$, respectively.

Further investigation of the system dynamics for a wider range of the parameters shows that changes of gain or loss rates at fixed large $\Gamma$ do not lead to any dramatic changes: while the predicted tendency $n\rightarrow2(\gamma-\kappa)/\Gamma$ becomes incorrect, $\rho(t)$ and $\phi(t)$ still display usual damped oscillations around the slowly changing values defined in the right-hand sides of Eqs.~(\ref{rho_lin}) and (\ref{phi_lin}). Typical trajectory of the pendulum on the sphere for this case is plotted in Fig.~\ref{fig_spheres}(b) for $\gamma=4\gamma^{\mbox{\footnotesize thr}}$ and $\Gamma/g=1$. As explained in the main text, the trajectories starting from the basin of attraction of the lower fixed point (which exists at large values of $\Gamma/g$) flow towards the unstable limit cycle, and then relax towards the stable lower focus which is located on the ``photon'' semi-sphere ($\rho>0$, $\phi>\pi$). However, the decrease of $\Gamma$ causes the complete change of this behavior.

\subsection{Inverted pendulum state}

As gain rate $\gamma/\gamma^{\mbox{\footnotesize thr}}$ increases at fixed $\Gamma/g$, or, equivalently, the saturation parameter $\Gamma$ decreases at fixed $\gamma$, the condensate population grows and hence the interparticle interactions strongly alter the dynamics.

\begin{figure}[t]
\renewcommand{\captionlabeldelim}{.}
\includegraphics[width=\columnwidth]{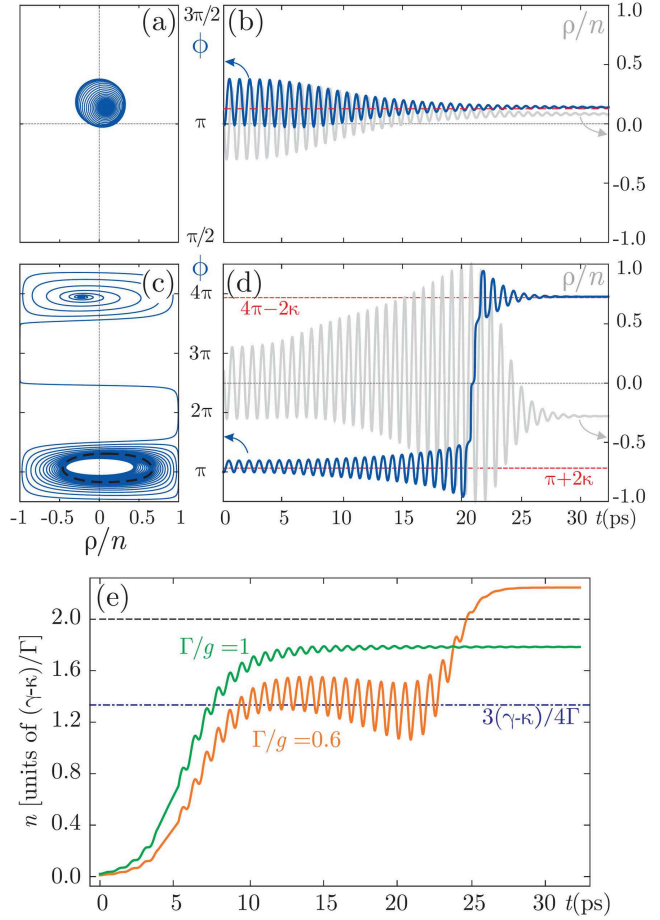}
\caption{\linespread{1}\small Numerical solutions of the evolution equations (2)--(4) of the main text for $\kappa=0.1$, $\gamma=0.2=2\gamma^{\mbox{\footnotesize thr}}$ and initial values $\rho(0)=0.35n$, $\phi(0)=\pi+2\kappa$. (a) Trajectory projection on the phase-plane $(\rho/n,\phi)$ for $\Gamma/g=1$, and (b) corresponding evolutions $\rho(t)$ (grey) and $\phi(t)$ (blue). The red dashed line indicates the focus coordinate $\phi_\infty^{\mbox{\tiny LP}}$ given approximately by (\ref{focus_LP}). (c) Phase-plane portrait projection for $\Gamma/g=0.6$ showing crowding of the trajectories in the area of the limit cycle projection (black dashed line) and consequent stabilization of the inverted pendulum state at $\phi\simeq4\pi-2\kappa$. (d) Corresponding evolutions  $\rho(t)$ (grey) and $\phi(t)$ (blue). The red dashed lines show $\phi_\infty^{\mbox{\tiny LP}}$ and $\phi_\infty^{\mbox{\tiny UP}}$ given by (\ref{focus_LP}) and (\ref{focus_UP}), respectively. (e) Total density $n$ against time for $\Gamma/g=1$ and $0.6$ as marked. When passing the saddle limit cycle, $n(t)$ oscillates with $\langle n\rangle\approx 4(\gamma-\kappa)/3\Gamma$ (see the saddle point \textit{(c)} of the averaged diagram, Fig.~2 in the main text). Other parameters same as used in Fig.~\ref{small-gamma}.}\label{SFig_1}
\end{figure}

The case of relaxation oscillations discussed in the previous section is shown in Fig.~\ref{SFig_1} in comparison with the ``inverted pendulum'' scenario. We fix the initial conditions $\rho(0)=0.35n$, $\phi(0)=\pi+2\kappa$ and pumping $\gamma=2\gamma^{\mbox{\footnotesize thr}}$, and change the non-linear loss rate $\Gamma$. While for the both cases the starting point lies inside the limit cycle projection on the phase portrait $(\rho,\phi)$, for $\Gamma/g=1$ (see Fig.~\ref{SFig_1}(a),(b)) it belongs to the basin of attraction of the lower focus, in contrast to the case of $\Gamma/g=0.6$ shown in Fig.~\ref{SFig_1}(c),(d) where the population imbalance oscillations amplify up to their maximum amplitude $\rho_m=n$. Note that when the amplitude is large, the analytical damping rates found in the Sec.~III are no longer correct, and the oscillations of the relative phase start to amplify as well. When $n(t)$ reaches its saddle point, it starts to oscillate around $\langle n\rangle\approx3(\gamma-\kappa)/4\Gamma$ (see Fig.~\ref{SFig_1}(e)), while the trajectories on the phase plane $(\rho,\phi)$ get crowded around the saddle limit cycle projection shown as the black dashed line in Fig.~\ref{SFig_1}(c). After leaving the limit cycle, the evolution line relaxes fast towards the upper stable focus (\ref{focus_UP}).

\begin{figure}[t]
\renewcommand{\captionlabeldelim}{.}
\includegraphics[width=\columnwidth]{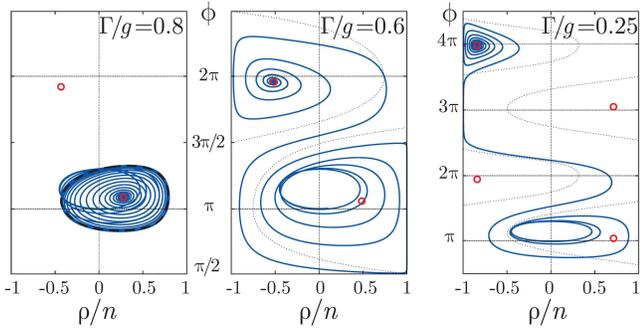}
\caption{\linespread{1}\small Phase-plane portraits (projections of 3D trajectories on the plane $(\rho/n,\phi)$) for $\kappa=0.1$, $\gamma=0.4=4\gamma^{\mbox{\footnotesize thr}}$, and $\Gamma/g$ as indicated on the panels. Initial conditions: $\rho(0)=0.1n$, $\phi(0)=\pi$. Other parameters are the same as used in Fig.~\ref{small-gamma}. The black dashed line marks the projection of the saddle limit cycle where it exists, the red circles indicate the fixed points. When the starting point belongs to the basin of attraction of the lower equilibrium, after reaching the limit cycle the trajectory winds up to the stable LP state (see left panel). Otherwise the system arrives at the ``inverted pendulum'' (UP) state. Thin dotted lines in the central and right panels schematically mark the corridor of infinite motion (see the text for more details).}
\label{SFig_2}
\end{figure}

In Fig.~\ref{SFig_2}, we increase the pumping $\gamma=4\gamma^{\mbox{\footnotesize thr}}$, and investigate again the system behavior for different values of the ratio $\Gamma/g$ at fixed initial conditions $\rho(0)=0.1n$, $\phi(0)=\pi$. Qualitatively, the dynamical regimes are the same as for small $\gamma$, although the dynamics is much faster, and the UP state is much more shifted towards the exciton state, in agreement with (\ref{focus_UP}). The corresponding evolution on the normalized sphere (the trajectory of the pendulum) is presented in Fig.~\ref{fig_spheres}(c).

However, in Fig.~\ref{SFig_2} we notice that with the decrease of $\Gamma$ and destabilization of the LP equilibrium, the transition to the UP state happens without population imbalance amplitude reaching its maximum possible value $n$. The smaller is $\Gamma$ compared to $g$, the smaller is the amplitude of oscillations at which the transition to the upper focus happens. One could understand this effect as follows. With higher pumping, the population of the condensate increases, and the blueshift value $g|\psi_X|^2$ adds more to the (negative) detuning between the photon and exciton modes. As we have shown in our previous work \cite{voronova_prl} for a conservative system without gain and dissipation, when detuning increases, there can be two regimes of internal oscillations: that of Rabi oscillations with the relative phase oscillating around $\pi$ (and trajectories on the phase plane orbiting the fixed points), and the ``internal Josephson'' regime characterized by the running relative phase, when the trajectory on the phase plane becomes reminiscent of the ones shown in the central and right panels of Fig.~\ref{SFig_2}, however infinite (not decaying towards the foci). Here, one could say that the transition from oscillations around the lower focus to oscillations around the upper focus happens via the ``internal Josephson'' regime. When the density increases (due to the pump increase or the losses decrease), the effective detuning grows, and the area of bounded motion (closed orbits) on the phase plane $(\rho,\phi)$ reduces, while the corridor of infinite motion (shown as thin dotted lines in Fig.~\ref{SFig_2}) widens. Hence the smaller amplitude of $\rho$ is needed to get into this area and consequently transit towards one of the attracting foci $\phi_\infty=2\pi k-2\kappa$ ($k\in\mathds{Z}$). The foci given by $\phi_\infty=\pi(2k+1)+2\kappa$ ($k\in\mathds{Z}$) are repulsive, as one can see in the right panel of Fig.~\ref{SFig_2}. For more details about ``internal Josephson'' regime of the running relative phase please see Ref.\cite{voronova_prl}.

\end{document}